# Upstream and Downstream AI Safety: Both on the Same River?

*John McDermid, Yan Jia, Ibrahim Habli*
Centre for Assuring Autonomy, University of York, UK

ABSTRACT

Traditional safety engineering assesses systems in their context of use, e.g. the operational design domain (road layout, speed limits, weather, etc.) for self-driving vehicles (including those using AI). We refer to this as "downstream" safety. In contrast, work on safety of frontier AI, e.g. large language models which can be further trained for downstream tasks, typically considers factors that are beyond specific application contexts, such as the ability of the model to evade human control, or to produce harmful content, e.g. how to make bombs. We refer to this as "upstream" safety. We outline the characteristics of both upstream and downstream safety frameworks then explore the extent to which the broad AI safety community can benefit from synergies between these frameworks. For example, can concepts such as common mode failures from downstream safety be used to help assess the strength of AI guardrails? Further, can the understanding of the capabilities and limitations of frontier AI be used to inform downstream safety analysis, e.g. where LLMs are fine-tuned to calculate voyage plans for autonomous vessels? The paper identifies some promising avenues to explore and outlines some challenges in achieving synergy, or a confluence, between upstream and downstream safety frameworks.

## Introduction

There is currently global interest in safety of artificial intelligence (AI), particularly safety of foundation or frontier AI models, also known as general purpose AI (GPAI). The trigger for this was the emergence of ChatGPT in late 2022 and the rapidly advancing capabilities of the ChatGPT family and other large language models (LLMs). Indeed, some are concerned that GPAI may, in time, evade human control and could pose existential risks [1, 2]. There are also more specific concerns, e.g. the ability of such models to assist "bad actors" in developing harmful capabilities, e.g. bioweapons. These concerns have led to international agreement on the need for controls and the emergence of frameworks for assessing and managing safety of such AI models [3]. These frameworks tend to focus on how the capabilities of the AI models are developed, e.g. training and fine-tuning, and the risk controls that can be introduced by the model developers; we refer to these as *upstream* AI safety models, as they address the development of general AI capabilities rather than the multifarious services provided using those models, e.g. through user-facing apps.

In contrast, safety engineering (in the traditional sense) is a much older discipline which originated around 1950 in response to the risks associated with space flight and some military capabilities. The early processes focused on failure, e.g. loss of structural integrity or thermal runaway of a chemical process. There have been major changes since these early days. First, software has become a critical component of such systems, with approaches to improving the integrity of software/minimise the likelihood of software-related failures originating in the 1980s [4, 5]. Second, the introduction of autonomy (often enabled by AI) has led to a focus on assessing the safety of the intended function (hitherto *assumed* to be safe) [6]. Throughout this evolution, safety engineering practice focused on the product in its operational context, for example the same humanoid robot in a care home or a factory poses quite different risks and requires different risk controls. We use the term *downstream* safety to reflect assessment in context, drawing on traditional safety engineering.

This paper explores the similarities and differences between upstream and downstream safety; we hope that by clarifying concepts and identifying commonalities of concerns and practices we can facilitate constructive dialogue. More ambitiously, we hope to identify ways of improving traceability between upstream and downstream safety; we believe this is fundamental for safety assurance, governance and transparency, and for allocating responsibility (or more positively for empowering the right parties to take responsibility for their roles and actions). In this sense, we hope to show whether the upstream and downstream safety models are on the same river – or at least there is a confluence in sight.

We start by presenting some of the relevant concepts of downstream safety and reinforce the value of analysing safety in context, as this determines the potential nature and extent of harms. We then consider upstream safety, including re-articulating some of the emerging concepts using downstream safety terminology. This leads on to a discussion of the potential for a confluence between the two approaches, what would be needed to achieve it, some views on future research directions and the implications for AI regulation.

## Downstream Safety

Downstream safety has traditionally focused on physical harm (to humans and sometimes the environment) [7]. It analyses systems or operations, starting from the concept of a hazard – a situation which could give rise to harm, e.g. wrong dosage rate for a syringe pump, then identifying failures that could cause the hazard (whether human or technical) down to "basic" faults, e.g. a valve sticking or a resistor failing open circuit. Controls are identified, e.g. redundancy, diversity and developing components to high levels of integrity to reduce the risk to an acceptable level. The notion of design integrity applies to physical components but is also widely used for software, with the controls involving increased rigour and depth of verification evidence as the consequences of failure/malfunction increase [8]. Approaches to achieving and assuring software integrity are often defined in standards, and these vary across application domain, e.g. aerospace [9] vs healthcare [10].

With the growing introduction of autonomous capabilities, e.g. in cars and unmanned aircraft, there has been a more explicit recognition of the safety of the intended function (SOTIF), i.e. assessing whether the system is safe if it operates as intended, rather than the traditional focus on failures (along with the implicit assumption that working as intended is safe). Work on SOTIF is most explicit in automotive, which had perhaps overlooked the issue, and a recent standard [11] sets out the need to reduce uncertainty when a system (perhaps including AI) is operating in a complex and only partially knowable environment [12]. Further, there is work on assurance of AI in several embedded systems domains, e.g. automotive[1] and aerospace, and some domain-agnostic approaches[2]. These are generally extensions of well-established safety engineering approaches, considering AI as part of a systems engineering lifecycle. Many of these approaches embrace a safety case, that is a structured argument supported by evidence intended to justify that a system is acceptably safe for a specific application in a defined operating environment; the safety case is likely to be communicated to a regulator and used as the basis of approvals [13, 14].

---

[1] ISO/PAS 8800 Road vehicles — Safety and artificial intelligence (to appear).
[2] Assurance of Machine Learning for Autonomous Systems (AMLAS), see: https://www.york.ac.uk/assuring-autonomy/guidance/amlas/

The "underlying science" behind the safety analyses and safety cases is modelling *causal dependencies between faults and failures* (and functional capabilities in the case of SOTIF). A range of analyses are employed, e.g. HAZard and OPerability Studies (HAZOP) [15] which is an exploratory method considering how potential deviations from intended functionality might contribute to hazards, and fault tree analysis (FTA) [16] which considers how individual subsystem or component failures might combine to cause a hazard. FTA works top-down, typically from a known hazard and deduces its causes; in contrast, failure modes and effects analysis (FMEA)[3] works bottom-up from known failure modes of sub-systems or components to investigate their system-level effect. Design is conducted so that no single point of failure (SPOF) will give rise to a hazard by mitigating all individual failures, e.g. using redundancy. These analyses are used to "drive" the design, typically producing derived safety requirements (DSRs) for additional functions to control known failure modes, for redundancy/diversity, and for the level of integrity required for critical components [18].

There are some specific concerns in analysing complex systems. First, identifying common mode failures (systems failing the same way at the same time, perhaps as they use the same software design) and common cause failures where the same underlying issue, e.g. flood or loss of cooling, leads to (near) simultaneous or highly correlated failures. Common cause failures can also be common mode and both can be SPOF. Second, there is a notion of "particular risks" where there are potentially far-reaching consequences of a particular (sub-) system failure mode; an uncontained aero engine failure which can cut power, hydraulic and fuel lines, as well as lose part of the aircraft propulsion is a paradigmatic example[4].

Downstream safety is highly contextual – indeed this is explicit in the definition of a safety case. The humanoid robot example mentioned above is indicative: in a care home the robot might need to interact closely with frail individuals whose movement might be unpredictable; risk controls will include force limitation when in contact with individuals and a focus on the integrity of the control software. In a factory the people at risk are likely to be maintenance staff, and the most important risk control may well be the ability to disable the autonomous capability before maintenance work starts (and to re-enable autonomy from a safe distance). Some standards, e.g. ISO 26262 for automotive systems [19], introduce the idea of a safety element out of context (SEooC) which can be analysed independent of use. In the authors' experience[5] this is difficult as the variation in usage can be significant and there is a need to analyse all possible usage scenarios. As an example, consider a flat (pancake) electric motor for an electric vehicle. There might be one per wheel, one at the back (on the axle) and one for each front wheel, or the usage may be for the rear wheels only (one on the axle or two motors, with one for each wheel). Dealing with the variety is more complex than dealing with an actual design, i.e. analysing in context.

---

[3] There is no definitive reference for FMEAs but there are some surveys of approaches, e.g. [17].
[4] See: https://www.atsb.gov.au/publications/investigation_reports/2010/aair/ao-2010-089 for an analysis of an uncontained engine failure on an Airbus A380.
[5] One of us was involved in advising on the electric vehicle example outlined here.

Downstream safety requires visibility of the design. Thus, much of the analysis work is done by developers but with independent scrutiny. Details vary by sector, but generally the level of independence increases with the potential for harm. In many sectors the safety case is started early and evolves through life and is used for a final independent check in support of regulatory approval prior to deployment. This model has worked well in many domains, e.g. aerospace and healthcare, including dealing with substantial changes in technology, e.g. introduction of composite materials in aircraft and the growing use of computer-based decision-support systems in hospitals. It is evolving to deal with AI, for example AI-based vision for SDVs, although it is proving hard to adapt at sufficient pace. Can the downstream model survive in the age of GPAI, or will the levee break, to stretch an analogy?

## Upstream Safety

Following the AI Safety Summits in the UK and Korea, 16 major GPAI developers agreed to publish their AI safety frameworks before the next summit (in France in February 2025). Our use of the term upstream safety is intended to encompass the work on these frameworks. There are several major differences to downstream frameworks.

First, the concerns go **wider** than physical harm (although they typically include it), for example addressing the ability for GPAI to help bad actors to develop bioweapons, the ability to manipulate political processes, e.g. through deepfake videos, and the ability to assist organisations in mounting cyber-attacks.

Second, the frameworks reflect much more **dynamic** processes. Downstream safety has, historically, focused on assuring safety in support of a deployment decision. This has been seen as the critical decision point, with lower levels of scrutiny applied to updates to the system. In contrast, the upstream frameworks must deal with much faster (perhaps DevOps) development models and the fact that the deployed models can be adapted by users[6].

Third, the focus is (largely) on the *capabilities* of the AI models themselves, with less focus on actual application and context of use (this is unsurprising, given the generality of the technology). In terms of the downstream model there is a focus on SOTIF.

The emerging frameworks vary, but there are commonalities, reflecting the National Institute of Standards and Technology (NIST) Assessing Risks and Impacts of AI (ARIA) framework[7] which focus on evaluation (evals) to inform model refinement and to identify the need to introduce risk controls. Typically the evals encompass:

1. Model testing including benchmarking on tasks intended to reflect real-world usage.
2. Red teaming using experts to stress test and to seek to find flaws or weaknesses.
3. Field-testing in conditions representative of operational use (but usually controlled).
4. Long-term impact assessments.

To our knowledge, no-one has proposed an "underlying safety science" for the upstream safety model. We suggest, in order to prompt discussion, that this might be characterised as *capability evaluation* (reflecting point 3 above) and *evolutionary thinking* (reflecting point 2 above and the iterative model refinement process supported/guided by evals).

---

[6] The dynamics are changing for downstream safety, but the pace is still likely to be much slower than with GPAI.
[7] See: https://ai-challenges.nist.gov/aria.

There is an interesting comparison here with what is often called "Safety-II" [20] in the "downstream world" which emphasises that actual (positive) safety happens, specifically in highly operational and open contexts, because of the capability (or capacity) of the system (mainly through intelligent agents, i.e. people) to adapt/evolve/adjust in response to (hard to predict) events. The Safety-II approach also emphasises resilience and how to recover following adverse events. Although not explored in more detail here, there would be merit in exploring whether there are lessons for GPAI safety from Safety-II.

We now consider some of the emerging frameworks and other related developments in terms of capabilities and risk controls, drawing out further differences and similarities between the upstream and downstream views to inform a discussion about what each community can learn from the other.

*Level of capability and capability evaluation*

We use two examples to illustrate the ways in which GPAI developers characterise and evaluate model capabilities. Google DeepMind has set out a frontier safety framework[8] which they aim to implement by early 2025 before the considered risks materialise. It uses the term critical capability levels (CCLs) defined as "thresholds at which models may pose heightened risk without additional mitigation". They identify CCLs in four risk domains:

- Autonomy – models capable of expanding their capacity by acquiring resources to run additional copies of themselves.
- Biosecurity – ability to develop known biothreats easily (non-expert users) or novel biothreats (expert users).
- Cybersecurity – ability to fully automate cyber-attacks opportunistically or on critical national infrastructure.
- ML R&D – ability of the models to significantly accelerate the development of AI capability (presumably most powerful if applied recursively).

In the downstream model's terminology these are not hazards or failure modes; they are perhaps best thought of as particular risks. ML R&D can be thought of as a (potential) common cause (enabler) for the other three risk domains (and more). Similarly, autonomy has the potential to be a common cause for many other harms, or it could be benign (aside from resource consumption), e.g. producing pictures of cats, or beneficial, e.g. searching for biomarkers for types of cancer. In the downstream world view, the benefit or risk cannot be assessed except in the application context.

Finally, Google DeepMind discuss evaluation to provide early warnings of potential violations of CCLs. This is intended, *inter alia*, to allow researchers time to develop new mitigations before the risks materialise – which is a form of evolutionary thinking.

OpenAI has also developed guidelines on AI Safety and their evolving position is reflected on their website[9]. Their approach includes a Preparedness Framework[10] which includes what they term catastrophic risks against which they evaluate models prior to release. The four catastrophic risks they identify, which can also be seen as capabilities, are:

---

[8] See: https://storage.googleapis.com/deepmind-media/DeepMind.com/Blog/introducing-the-frontier-safety-framework/fsf-technical-report.pdf
[9] See: https://openai.com/safety/
[10] See: https://cdn.openai.com/openai-preparedness-framework-beta.pdf

- Cybersecurity – the use of a model to disrupt confidentiality, integrity and/or availability of computer systems (this is often referred to as the CIA triad).
- CBRN Chemical Biological Radiological and Nuclear (risks) – ability to assist in creation of CBRN threats.
- Persuasion – convincing people to change their beliefs or to act on model-generated content[11].
- Model autonomy – the ability to enable malign users of the model to evade control and ultimately for the model to do so on its own (in other documents referred to as autonomous replication and adaptation (ARA)).

OpenAI's evaluations include red-teaming and safety evaluations to identify the need for further risk controls, resulting in the production of system cards which identify the level of risk against the preparedness framework. The approach is different in detail to Google DeepMind's, but there are similarities including the range of concerns and the emphasis on evals and red teaming. Thus, rather than consider their framework in detail, we consider a recently announced development which may be of wider significance – the o1 model[12].

In their announcement, OpenAI explain that they use reinforcement learning to train o1 to undertake "chain of thought" (CoT) reasoning and use it to improve model performance and to give a way of summarising the model's "thinking". They evidence performance that exceeds previous models, e.g. Chat GPT4o, across a range of tasks including mathematical challenges and computer programming. They also consider its role in safety.

OpenAI show that o1 does much better than ChatGPT4o on jailbreaking evaluations, and is better on other criteria, e.g. handling harmful prompts. This is set out in a system card[13] which shows low or medium risks against each element in their preparedness framework. More significantly, by using the CoT in red teaming and evaluations they observed novel and interesting instances of reward hacking[14] thus there is *prima facie* evidence of the value of CoT reasoning in evaluating safety. More generally, the CoT reasoning can be thought of as a form of explainability giving a level of visibility to the models.

*Risk controls*

There seems to be more variation in the definition of risk controls than there are in the levels of capabilities and evaluation strategy.

Google DeepMind identify two classes of mitigation: security and deployment controls. The security controls focus on preventing exfiltration of model weights as this would allow others to replicate the models[15] and potentially to exploit the full capability of the models.

The deployment controls can be summarised as follows:

0. Safety finetuning of models and use of filters.
1. Full suite of prevailing industry safeguards plus periodic red teaming to assess the adequacy of the mitigations.

---

[11] There are many possible adverse effects, but one which is topical is disruption of election processes.
[12] See: https://openai.com/index/learning-to-reason-with-llms/
[13] See: https://cdn.openai.com/o1-system-card.pdf
[14] See page 16 in the system card (the previous footnote).
[15] We assume that the underlying concern is that the exported weights would contain enough information to allow the structure of the model to be inferred and hence reproduced.

2. Safety case to keep numbers of incidents below a prescribed level with red team validation pre-deployment.
3. Prevention of access to the model capabilities (an open research problem).

The description of level 2 doesn't use the downstream terminology for safety cases but implicitly it includes goals (what is to be demonstrated), strategies for decomposing goals into subgoals, and evidence to show that the goals have been met reflecting the standard structure of a safety case[16]. The underlying intent of level 2 deserves some scrutiny. Given the severity of the harms associated with the higher CCLs any number of incidents above zero would seem to be highly undesirable. One way of rationalising this would be to construct the safety case to include the mitigation measures, e.g. guardrails, and to identify "incidents" as triggering of the mitigations. Assuming there are multiple mitigations for the more severe outcomes, then setting targets for incidents in terms of mitigation measures should still result in acceptable risk, as several mitigations need to fail (or be "breached") to allow adverse effects. This might also link to Safety-II – the mitigations acting to prevent harm are part of the system "working right" and thus should be reinforced.

OpenAI focus on training of models and the use of guardrails providing a hierarchy of controls which aligns with the use of classical downstream representations of safety controls, such as bow tie models[17]. The training reflects a Model Spec[18] which includes:

- Objectives including benefitting humanity, reflecting social norms and complying with applicable laws.
- Rules including protecting people's privacy and respecting creators' rights.
- Default behaviours which are as much about the way developers and users conduct themselves as the AI models themselves.

The Model Spec is mainly intended to be used to guide developers. OpenAI also refer to sharing real-world feedback which is similar to the NIST ARIA long-term impact assessment.

Some recent work on safety of GPAI builds on the concept of safety cases, adapted from the downstream safety world. One approach identifies "building block arguments" [19], see Figure 1 below, with the expectation that the arguments used will vary as the models become more powerful. This approach strongly links capabilities and controls.

The building block arguments are, in summary:

- Inability – the model cannot cause unacceptable outcomes in any realistic setting.
- Control – unacceptable outcomes are prevented by existing control measures.
- Trustworthiness – even though the models could cause catastrophic outcomes, they won't as they robustly behave as intended (there is an obvious analogy with SOTIF).
- Deference – an AI advisor asserts that the AI model doesn't pose a catastrophic risk (and the AI advisors are at least as credible as human decision-makers).

As GPAI models will evolve rapidly the safety case will need to be dynamic [22].

---

[16] For example, the goal structuring notation (GSN), see: https://scsc.uk/r141C:1
[17] See: [21] for an example for the use of bow ties for representing the role and inter-relationships of risk controls for AI (the term barriers is used) and some of the other downstream safety concepts used here.
[18] See: https://openai.com/index/introducing-the-model-spec/
[19] Se,e: https://arxiv.org/pdf/2403.10462

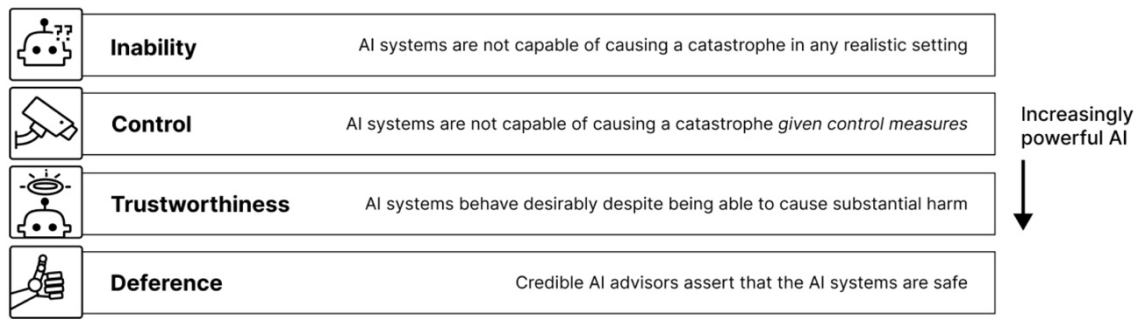

Figure 1: Proposed Safety Case Arguments (see[17])

As an illustration, the argument supporting inability includes red teaming and fine tuning of models, and an assessment of the capability of the red team. It is unclear how strong such an argument can be without full visibility and understanding of the models which seems unrealistic even for simple GPAI models which will have millions of weights (if not billions).

Deference means that "AI is guarding AI". In downstream terms, this seems to be prone to common mode failures unless a different model with different training data can be used as the advisor. In computer science terms, this seems akin to solving the halting problem, so the proposal seems highly questionable.

Finally, they suggest a dialectic approach where developers produce a safety case and a red team produces a risk case, both of which are presented to regulators. Similar approaches have been suggested in the downstream safety community. Given the uncertainty around advanced AI models the dialectic approach might have considerable merit in allowing AI safety claims to be independently assessed and challenged.

*Observations*

Despite the differences in the details the approaches reviewed above have some common elements, viz: a focus on evals (e.g. testing against a range of benchmarks), use of red teaming and an emphasis on fine-tuning to achieve model alignment. They can perhaps be thought of as refinements of NIST's ARIA model. Also, they share a concern about model transparency (aka visibility or explainability) but the approaches, perhaps excepting the use of CoT reasoning with OpenAI's o1, do not consider the issues of achieving transparency with large-scale GPAI models[20].

From a downstream safety perspective, where unsafe failure rate targets might be one in a million operations/hours or less, the evaluations, e.g. showing jailbreak "successes" in the 10s of percent, seem a very long way away from acceptable levels of risk. Or are they? That depends on the context of use, which is largely lacking in these upstream frameworks.

# Comparison and Analysis

We first set out a comparison between the upstream and downstream models, then consider what each might learn from the other. Table 1 summarises our views on the core characteristics of upstream and downstream safety, including identifying similarities. We briefly discuss the key points in the table, then consider what it would mean for upstream and downstream safety to be "on the same river".

---

[20] And it is not clear from the paper how accurately the CoT reflects what the model is actually doing.

*Table 1: Comparison of Upstream and Downstream Safety*

| Viewpoint | Underlying Science | Risk Identification | Risk Controls | Similarities | Insights |
|---|---|---|---|---|---|
| Downstream Safety | Mainly causal dependencies thinking (but recently re-emphasised concerns with intended function). Safety-II provides a complementary framework [20]. | Analysing risk through identification of hazards, in the application context, their potential causes and effects, to make the risks concrete and visible thus amenable to control. | Controls are mainly architectural, e.g. designing for safety via redundancy and diversity, protecting against failures, and through development of critical components to high levels of integrity[21]. | Concern with safety of the intended function (SOTIF) analogous to GPAI capabilities, and analysis of Safety Element out of Context (SEooC) similar in concept to evals. | Value of assessing risk and safety in the context of use. The importance of architecture in controlling risk, including avoiding SPOF. |
| Upstream Safety | Capability achievement and evolutionary thinking (shape the evolution of the models to ensure safety or alignment with human values). | Analysing risk through level of capability (in general terms and in specific terms, e.g. cybersecurity) and its potential impact on society or humanity if unchecked. | Heavy emphasis on model testing and evaluation, e.g. red teaming, together with monitoring and feedback, leading to improvement of the model. In addition, use of safeguards, e.g. input filters, to control usage of the models. | As above. In addition, some work on upstream safety on concrete problems in AI [23] does consider what the downstream safety world would view as defining "physics of failure" (see below). Further, model improvement is similar to developing high integrity components. | The value of understanding the capability of the models. The emphasis on evaluation and the use of ongoing evaluation as the models evolve. |

---

[21] There are often also procedural safeguards reflecting the need for human involvement, e.g. to provide resilience.

*Insights and challenges for upstream safety*

There are three key take-aways, or insights, from upstream safety assurance:

1. Risk is viewed as an aspect of capability and is generally assessed independent of context.
2. Confidence in the models comes from post-development assessment (evals) and the subsequent refinement of the models based on adverse findings in the evals.
3. There is nascent work on understanding failure modes of GPAI, which might be informative for downstream safety analysis and assurance.

The general challenge is coping with the capabilities and pace of change of GPAI models.

*Insights and challenges for downstream safety assurance*

There are four key take-aways, or insights, from downstream safety assurance:

1. The importance of analysing risk in context, including identifying hazards which can be viewed as the proximate cause of harm.
2. The need to identify failure modes, both top down and bottom up, and to understand how these can combine to give rise to hazards (and harm).
3. The need to identify "safety architectures" employing redundancy and diversity, but also the requirement for high integrity elements in the architecture to achieve safety.
4. The need to identify and understand common mode or common cause failures, and potential SPOF, which might undermine the safety architecture.

In terms of challenges, how can downstream safety cope with GPAI which is beyond the ability of established analysis methods to assess?

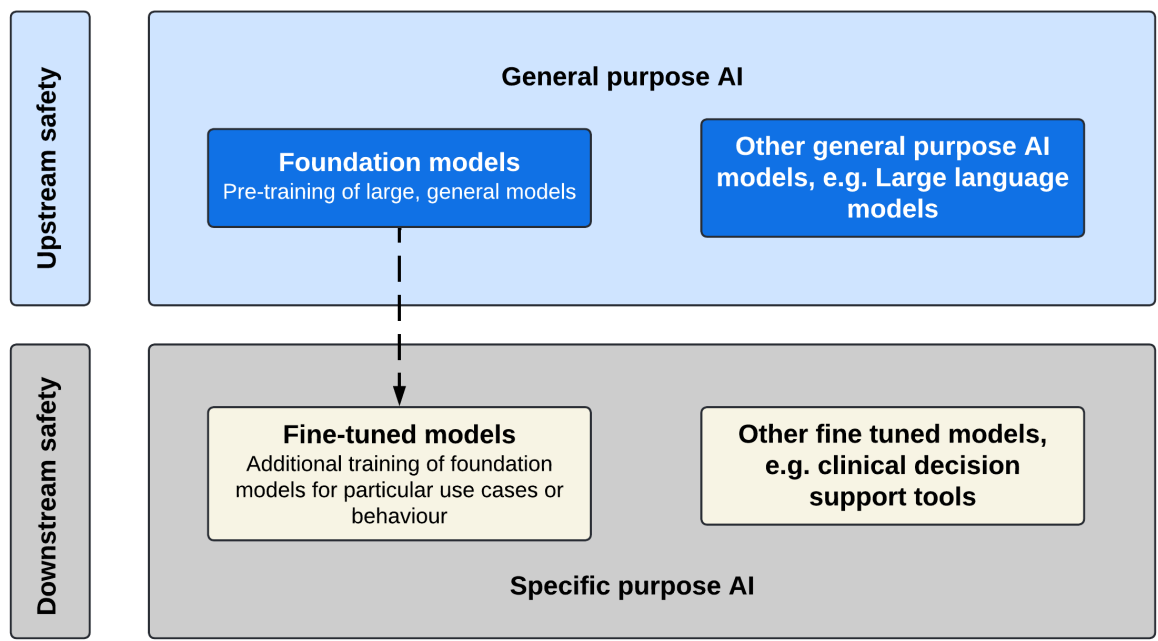

Figure 2: GPAI, Specific Purpose AI, Upstream and Downstream Safety

*The Same River or a Confluence?*

End-user applications of AI can arise in different ways, as shown in Figure 2. First, specific-purpose AI, e.g. for an SDV or for clinical decision-support (CDS) can be derived from GPAI, including frontier models. Second, such applications can be developed using more traditional AI, e.g. neural networks or reinforcement learning. Third, GPAI can be used "directly" – for example LLMs can be downloaded by end users and applied on many tasks across a wide range of application areas. Only in the first case is it important to understand how upstream safety frameworks might feed into downstream frameworks.

In a "downstream world view" the ideal would be that the upstream safety analysis provides information, e.g. on failure modes, that can be used in downstream analysis. Research on AI failure modes, e.g. identifying Concrete Problems in AI Safety [23], discusses issues such as distributional shift and reward hacking. In the downstream world view this is akin to physics of failure – a general *means* of causing failures, not a specific failure mode and we will call them "GPAI deviation classes"[22], adapting the term used in HAZOP. Thus, the linkage between the upstream and downstream models reflects ways in which these deviation classes could be manifest in the application context. For example, would the distributional shift between France and Germany's road infrastructure lead to unsafe trajectories for the SDV? Would the distributional shift between the USA and UK do so? This would have to be made more specific – for example, are there different road signs in France and Germany which would contribute to risks? Where the signs are the same, do they have the same meaning and the same sizes so that detection distances can remain the same? And so on.

Identifying hazards often uses guidewords as prompts, e.g. HAZOP considers on flows:

- Omission (not provided when intended).
- Commission (provided when not intended).
- Too much (flow, pressure, etc.)
- Other than (wrong substance, etc.)

When conducting HAZOP, consideration is given to whether such deviations could arise from known failure modes of the system and its components. HAZOP has been adapted for software and for AI, e.g. the perception elements of SDVs [24]. To extend this to GPAI and to provide the link between upstream and downstream safety would seem to require a more refined version of GPAI deviation classes, or a hybrid which allows the very general concepts, e.g. of reward hacking, to be interpreted in context. This is a research issue, but one that would help with identifying adverse effects of GPAI in specific application contexts.

Alternatively, it may be more effective to use the results of upstream safety analyses – model testing, red teaming, and field testing to identify deviation classes for downstream analysis. And there may be merit in refining the upstream analysis methods to characterise the identified deviation classes for downstream use. However, we cannot ignore other issues such as exfiltration of model weights. In the terminology of the downstream world, these could be viewed as particular risks – concerns that might have a much wider impact than the narrower deviation classes used in HAZOP[23]. Thus, there is a need to develop methods for exploring the potential impact of such risks in a systematic way. Given the complexity of the systems being considered this is likely to require mechanisation.

---

[22] At this stage this is speculative. 'Deviation' works, up to a point, at least for ''conventional systems', because of the tight scope of the system. We can more easily say what the system is deviating from in specific, meaningful and actionable ways. How this translates to GPAI would require more work at the conceptual and empirical levels, i.e. should be part of an overall GPAI safety research agenda.

[23] Similar to concerns with cyber-security weaknesses/vulnerabilities for conventional software.

Having identified adverse effects, it is desirable to prevent them, make them less likely, or to mitigate them (in the worst case mitigate their consequences). These come together in the system architecture – what combination of controls will be used, in what configuration, to reduce the identified risks to an acceptable level (in so far as it can be assessed)? The downstream world concerns itself with "safety architectures" with redundancy and diversity to address known failure modes. Addressing this is a research issue, but there is some initial exploration for "conventional AI" drawing on well-established fault-tolerance principles in aviation [25]. Another principle from downstream safety engineering is to use a hierarchy of controls, and not to rely on a single control; this concept is explored in a companion paper[24] which sets out a proposal on how to define layers of protection (hierarchies of control) for GPAI. Again, this would need further research, but it does seem to be "on the agenda" of some of the GPAI developers, e.g. OpenAI.

When defining architectures, exploration of potential common mode or common cause failures modes will consider the design and development process as well as the system itself. For GPAI these process issues will be critical; the most obvious concern is training data – two "different" models trained on the same data will likely have similar limitations, if not failure modes. Also, if the same underlying GPAI model is used, but it is adapted through fine-tuning or use of Retrieval-Augmented Generation (RAG) architectures to provide risk controls we would need to explore whether the limitations of the underlying GPAI model could affect both in the same way. This seems highly relevant to the notion of "deference" in the work on safety cases for advanced AI referenced above; it is said to be the most powerful form of safety argument, but it would also seem to be the most prone to common mode failures, particularly if the model learnt how to "game" its evaluation.

We now return to the subject of safety cases, which are widely used in safety-critical domains, including as a regulatory tool, and more recently for AI-enabled systems (downstream safety). Figure 3 sketches an overall AI safety case. The safety argument structure represented using the patterns and modular features of the Goal Structuring Notation (GSN)[25], integrates downstream and upstream safety assurance. It incorporates the following sub-arguments:

- **AI Ethics Argument:** Here, safety is one of several ethical principles (including fairness, privacy, and transparency) that must be assured for the use of AI. Trade-offs between risks and benefits will often be inevitable and must be justified and accepted by affected stakeholders or their representatives (e.g., regulators). Examples of AI ethics argument patterns can be found in [26, 27].
- **AI System Safety Argument:** This covers the wider system in which an AI model is deployed, which could be physical (e.g., an autonomous vehicle) or procedural (e.g., a sentencing system). The argument considers claims and assumptions about relevant hazards and risks, and evidence for the suitability of system-level risk controls including redundancy, diversity, monitoring and meaningful human control/ oversight. Example guidance for system-level AI assurance can be found in [28, 29]].

---

[24] AI Guardrails: Concepts, Models and Methods, John McDermid, Kester Clegg, Yan Jia, Ibrahim Habli.
[25] Briefly, in GSN, each 'folder' represents a module containing an argument; a solid ball is used for multiple instantiations; a hollow ball indicates 'optional' instantiation; a solid diamond reflects a choice. For more details, see the GSN Standard: https://scsc.uk/r141C:1

- **Purpose-specific AI Model Safety Argument:** This represents claims and evidence for AI models trained and tested for a specific purpose, such as identifying pedestrians using a CNN or recommending a treatment using RL. AMLAS provides practical guidance on purpose-specific AI safety cases [30].
- **General-Purpose AI Model Safety Argument:** This represents safety claims about capability-specific risks and guardrails, and the supporting evidence, particularly from evals and red teaming. Initial work in this area is represented by [31, 32, 33].

It is important to note that RAG architectures are increasingly utilised to improve the accuracy of GPAI models for specific purposes. In such cases, the safe use of RAGs should be justified within the 'Purpose-specific AI Model Safety Argument'.

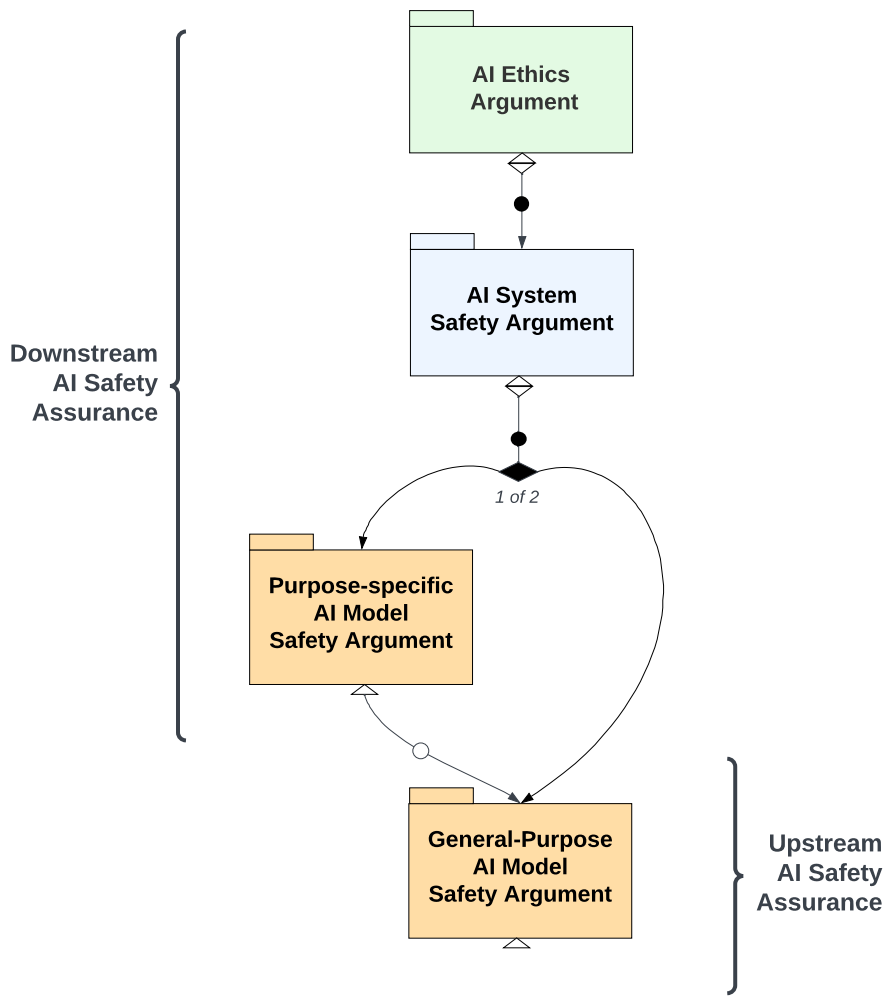

Figure 3: A Modular AI Safety Case Representation [34]

*Regulatory challenges in upstream safety assurance*

In regulatory terms, authorities typically have defined domains of responsibility, e.g. the Maritime and Coastguard Agency (MCA) in the UK and the US Coastguard Agency in the USA cover commercial shipping. They, and their counterparts in other nations, would be responsible for assuring the safety of GPAI deployments in the maritime context, e.g. use of AI for voyage planning and fuel optimisation[26]. They neither have the vires nor the capacity to consider risks beyond their domains. However, countries are establishing national AI regulators with a focus on GPAI. Based on Figure 2, it would make sense that such regulators focus on the "direct" uses of such models, e.g. using GPAI to influence the outcomes of elections, and act as knowledge sources for the domain-based regulators who must consider specific purpose systems, employing GPAI, in their sphere of concern.

---
[26] https://fintech.global/2024/05/29/orca-ai-secures-23m-investment-boost-after-north-standard-collaboration/

Moreover, such regulators could focus on the "particular risks" as we have termed them and their potential impact across domains – for example, what happens if the booking systems for all transport modalities are affected in the same way at the same time – knowing that "narrower" concerns would be addressed by the domain-based regulators. Of course, in an ideal world, if domain-based regulators identified issues of wider concern, i.e. with potential impacts well beyond their domains of responsibility, then they should report them to national AI regulators to bring them to the attention of the wider national and international community[27].

Seen like this, we would suggest that upstream AI safety can be seen as a tributary of downstream safety, but that it also remains a river in its own right. If the existing domain-based regulators address GPAI within their areas of responsibility, but collaborate with national AI regulators as outlined above, then the two streams might merge again through the regulatory ecosystem[28].

Conclusions

We have explored the world views of the AI community when discussing AI safety and that of the more traditional safety community when considering the impact of AI and GPAI; in doing so we have sought to find commonalities as well as differences. We argue that downstream safety is essential to understand the potential adverse impact of GPAI behaviour and failures *in context*. We are not alone. A recent Chinese report on their AI safety governance framework[29] also distinguishes inherent GPAI safety risks (upstream) and safety risks in GPAI applications (downstream). Thus, we conclude that the downstream model of safety is still relevant in the GPAI world, but we also believe it needs to (continue to) adapt to survive.

To this end, we sought to identify whether there is a confluence between upstream and downstream safety including identifying some potentially unifying terminology. More importantly, we have identified some possible ways that analyses of GPAI models could be used to inform the safety analysis of their applications. In essence this would be through providing guidance on classes of deviation from intent, and their causes, that can arise from GPAI models to inform downstream analysis, e.g. through adaptations of methods such as HAZOP. There is also a need to identify what might be seen as "particular risks" in the applications of the GPAI models. This requires collaboration between the two communities as knowledge and understanding of each is needed to make substantive progress. We see a research programme on adapting classical safety methods to reflect the characteristics of GPAI as being vital to progress.

---

[27] We assume that the national regulators will operate as a collaborative network.
[28] This would be a form of anastomosing river, which is, perhaps, taking the analogy too far.
[29] See: https://www.dlapiper.com/en-gb/insights/publications/2024/09/china-releases-ai-safety-governance-framework for a discussion in English.

In practice, a key issue will also be mechanisation. Classical safety work is largely carried out by hand; this is intractable at the scale of GPAI applications and the pace of change. Thus, for example, any work on deviation classes needs to be supported by tooling that allows experiments, e.g. injecting simulated deviations into models, and perhaps by using GPAI itself to search for and characterise GPAI deviation classes.

We have also discussed fault-tolerant system architectures and the ability to protect against (or control) failures through use of redundancy and diversity and the need to avoid common modes of failure. This needs to be extended into resilience – how systems can recover from failures – and incident investigation (to avoid recurrence of problems). This would build on and extend the notions of monitoring which are in most GPAI safety frameworks. Again, collaboration between the AI safety and traditional safety communities is needed here.

Assessing safety and regulating GPAI is challenging due to the complexity of the models and the uncertainties around their behaviour. There is no simple solution to this problem and the philosophy must involve reducing uncertainty prior to deployment and monitoring in operation to detect and respond to "the ones that got away". However, one potentially attractive pre-deployment activity is to use a dialectic approach – developers/would be deployers of GPAI present a safety case, and the red team presents counterarguments (a risk case) although this will mean changes in practice for existing (domain-based) regulators.

We hope that our discussion will have a further benefit in that it identifies the way in which established regulators in specific domains who are confronted with uses of GPAI can work with emerging national AI safety regulators, thus helping to shape the evolving AI safety regulation ecosystem.